# Multi-Channel SQUID System for MEG and Ultra-Low-Field MRI

Vadim S. Zotev, Andrei N. Matlachov, Petr L. Volegov, Henrik J. Sandin, Michelle A. Espy, John C. Mosher, Algis V. Urbaitis, Shaun G. Newman and Robert H. Kraus, Jr.

*Abstract*—A seven-channel system capable of performing both magnetoencephalography (MEG) and ultra-low-field magnetic resonance imaging (ULF MRI) is described. The system consists of seven second-order SQUID gradiometers with 37 mm diameter and 60 mm baseline, having magnetic field resolution of 1.2 - 2.8 fT/√Hz. It also includes four sets of coils for 2-D Fourier imaging with pre-polarization. The system's MEG performance was demonstrated by measurements of auditory evoked response. The system was also used to obtain a multi-channel 2-D image of a whole human hand at the measurement field of 46 microtesla with 3 by 3 mm resolution.

*Index Terms*— biomagnetism, MEG, ULF MRI, SQUID.

## I. Introduction

Magnetoencephalography (MEG) has been extensively used for real-time non-invasive studies of human brain function [1]. Its clinical application, however, has been impeded by the need to superpose the MEG-localized neuronal activity with anatomical information obtained using a separate high-field magnetic resonance imaging (MRI). Acquiring functional (MEG) and anatomical (MRI) data with two separate instruments leads to significant (typically ~5 mm or more) co-registration errors [2].

Magnetic resonance imaging at ultra-low fields (ULF MRI) is a promising new method for structural imaging at measurement fields in the microtesla range [3]-[6]. In this method, nuclear spin population is pre-polarized [7] by a relatively strong (up to 0.1 T) magnetic field, and spin precession is detected at an ultra-low (~100 µT) measurement field after the pre-polarizing field is removed [3]-[6]. The ULF MRI signals are measured by superconducting quantum interference device (SQUID) sensors [8], that can reliably operate at ultra-low fields.

Because the MEG array of SQUID sensors can be used for detection of ULF MRI, the two techniques are compatible and can be combined in a single instrument [6]. Moreover, it has been demonstrated that MEG signals and ultra-low-field nuclear magnetic resonance signals from the human brain can be acquired *simultaneously* by the same SQUID [9]. Thus, the combination of MEG and ULF MRI will allow simultaneous functional and anatomical imaging of the human brain. This will enhance neuroimaging by reducing or eliminating sources of error associated with co-registration. Moreover, because ULF MRI is free from distortions caused by susceptibility variations, anatomical images obtained at ultra-low fields can be used to correct image distortions in high-field MRI. In addition, the use of large MEG-style SQUID sensor arrays will allow faster imaging based on multi-sensor parallel imaging techniques.

In this paper, we describe a seven-channel SQUID system designed for acquiring both MEG and ULF MRI data. We demonstrate the system's MEG efficiency and present results of multi-channel ULF MRI of a human hand.

## II. Instrumentation

### A. Measurement system

The system we have built for measurements of MEG and ULF MRI signals consists of seven axial second-order gradiometers with SQUID sensors (Fig. 1). The gradiometers have 37 mm diameter and 60 mm baseline. They are hand-wound on G-10 fiberglass formers using 0.127 mm niobium wire. The gradiometers are positioned in parallel (one in the middle and six others surrounding it in a circle as shown in Fig. 1, right) with 45 mm separations between coil centers.

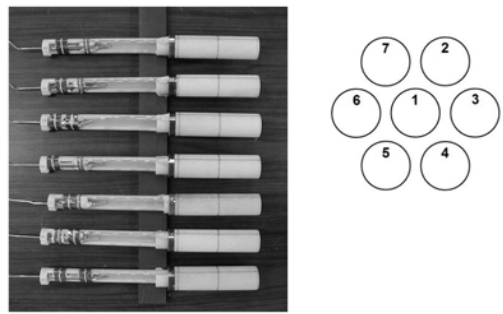

Fig. 1. Seven second-order gradiometers with SQUID sensors (left), and their positions inside the system (right).

Each SQUID sensor is equipped with a cryoswitch [10]. The cryoswitch, included in the superconducting input circuit

Manuscript received August 28, 2006. Development of the instrumentation was supported by U.S. Department of Energy, Office of Biological and Environmental Research, and by Los Alamos National Laboratory. The reported study was conducted in compliance with the regulations of the Los Alamos National Laboratory Institutional Review Board for research on human subjects and informed consent was obtained.

All the authors are with Los Alamos National Laboratory, Group of Applied Modern Physics P-21, MS D454, Los Alamos, NM 87545, USA (corresponding author V. S. Zotev, phone: 505-665-8460; fax: 505-665-4507; e-mail: vzotev@lanl.gov).

between the SQUID and the gradiometer, becomes resistive (50 ohm) when control current is run through its heater. The cryoswitch is activated during the pre-polarization in ULF MRI experiments, protecting the SQUID from transients caused by fast switching of the strong pre-polarizing field. The typical switching time is 5 µs. Both the SQUID and the cryoswitch are mounted inside a lead shield 12 cm above the gradiometer. The seven channels are installed inside a fiberglass liquid He dewar [11]. The system is operated inside a magnetically shielded room, and all equipment in the room is connected to the outside electronics via fiber optic cables.

Noise spectra of the seven channels are exhibited in Fig. 2. The magnetic field noise spectral density is ~1.2 fT/√Hz at 1 kHz for Channel 1, and 2.5-2.8 fT/√Hz at 1 kHz for the surrounding channels due to the dewar noise. The noise spectra remain essentially flat down to ~3 Hz. These data suggest that the system is well suited for MEG measurements.

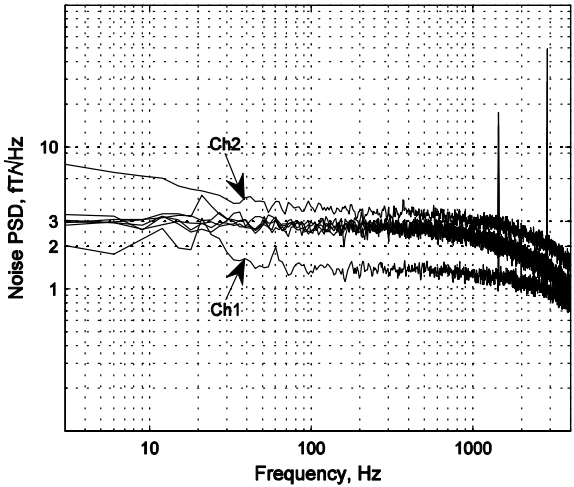

Fig. 2. Magnetic field noise spectral densities for the seven channels. The noise is lowest for Channel 1, and highest for Channel 2.

### B. Coil system

The coil system for 2-D ULF MRI experiments, depicted schematically in Fig. 3, consists of four sets of coils. A pair of round 120 cm Helmholtz coils provides the ultra-low measurement field $B_m$ along the Z axis. The field strength is 61 µT at 1 A. The longitudinal gradient $G_z = dB_z/dz$ is created by two 80 cm square Maxwell coils. A specially designed set of eight rectangular coils on two parallel 48 cm x 96 cm frames with 100 cm spacing provides the transverse gradient $G_x = dB_z/dx$. The magnitudes of $G_z$ and $G_x$ gradients are 120 µT/m and 80 µT/m at 1 A, respectively.

The pre-polarizing field $B_p$ is perpendicular to the measurement field $B_m$. It is created by a pair of thick round coils (25 cm outside diameter, 5 cm thickness), positioned around the dewar in a quasi-Helmholtz fashion with 23 cm spacing. This configuration allows pre-polarization of a large volume while minimizing response of the shielded room to strong $B_p$ pulses. This response is due to eddy currents in the room's aluminum plates, and has the characteristic time scale of up to 100 ms. The $B_p$ coils are made using litz wire. The strength of the pre-polarizing field is 10 mT at 12 A.

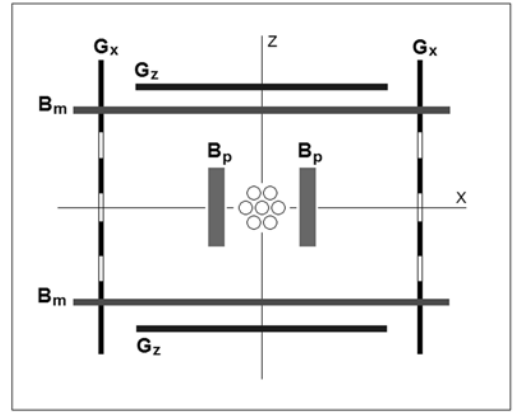

Fig. 3. Schematic of the coil system for 2-D ULF MRI measurements.

### C. Imaging procedure

The imaging procedure used in our ULF MRI experiments is illustrated in Fig. 4. It includes the pre-polarization stage, followed by the conventional 2-D gradient-echo sequence.

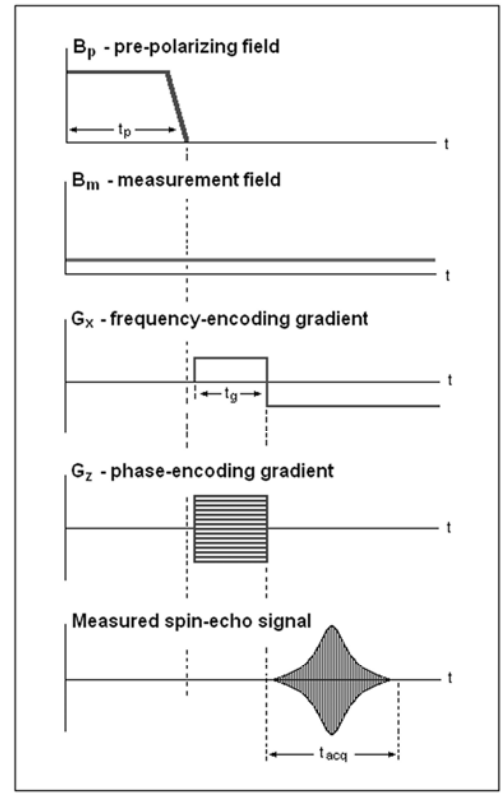

Fig. 4. Gradient-echo imaging sequence in 2-D ULF MRI experiments.

In the case of the human hand imaging, described in Section III.B, the following parameters were used. The hand was pre-polarized at field $B_p$ = 10 mT for 0.5 s. The pre-polarizing field was then removed non-adiabatically (in 0.5 ms), and spin precession took place at a much lower measurement field $B_m$ ~ 46 µT, with the Larmor frequency of ~1950 Hz. The precession was encoded during time $t_g$ = 40 ms by two gradients with limiting values of ±94 µT/m (±40 Hz/cm). The gradient coils were disconnected during the pre-polarization, and re-connected 5 ms after the removal of $B_p$ to avoid transients. This was done by means of optically isolated solid-state relays. The phase-encoding gradient $G_z$ was changed in 4.9 µT/m (2.1 Hz/cm) steps. Reversal of the frequency-encoding gradient $G_x$ created a spin-echo signal that was measured for $t_{acq}$ = 80 ms. The cryoswitch and the SQUID electronics reset function, activated prior to the pre-polarization, were turned off 3 ms and 5 ms after the gradient reversal, respectively. The described sequence provided 3 mm x 3 mm imaging resolution.

## III. RESULTS

### A. MEG measurements

In order to test efficiency of the seven-channel system in detection of MEG signals, we carried out measurements of auditory evoked magnetic field (auditory MEG). A train of clicks was used as auditory stimulus. The individual clicks were 1.2 ms long, and followed with 14 ms intervals. Duration of the click train was 100 ms with 50 ms pre-stimulus interval between the start of data acquisition and the stimulus onset. The subject's head was positioned under the dewar so that Channels 1, 3, and 6 were above the Sylvian fissure.

Results of the auditory MEG experiment are exhibited in Fig. 5. They show peaks at 40, 100, and 200 ms after the stimulus onset, characteristic of the auditory MEG [1]. The opposite signs of the signals from Channels 1, 2, 3, 4 and Channels 5, 6, 7 are consistent with the expected orientation of the equivalent current dipole perpendicular to the Sylvian fissure [1]. These results show that the seven-channel system can be successfully used for MEG measurements.

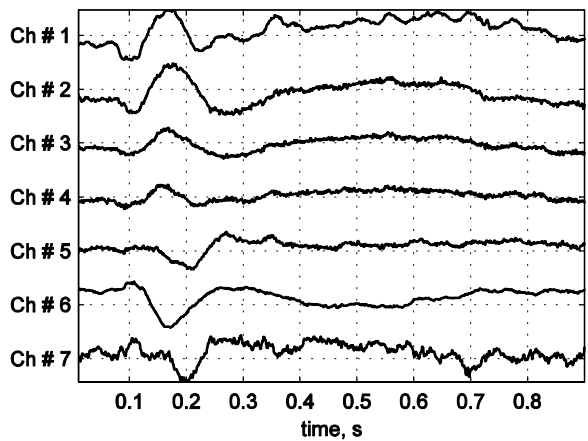

Fig. 5. Auditory MEG signals, measured by the seven channels.

### B. Human hand imaging

The described seven-channel system has been developed with the ultimate goal of imaging parts of the human brain. To evaluate the system's current ULF MRI performance, we obtained a 2-D image of a human hand. The hand's position during the experiment is shown in Fig. 6. The mean relaxation time $T_1$ for the hand was measured to be ~120 ms. Imaging was performed as described in Section II.C.

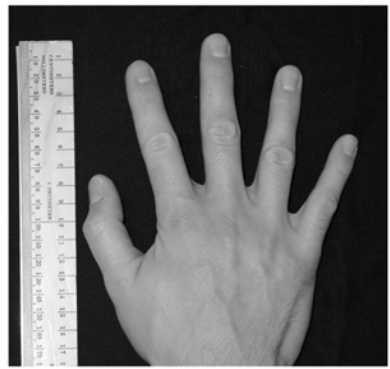

Fig. 6. The human subject's hand as positioned during imaging.

The composite image of the hand is shown in Fig. 7. It was obtained as a square root of the sum of squares of images acquired by the seven individual channels. Each individual image was a result of 2-D Fourier transform of the spin-echo signals measured for all selected values of the phase-encoding gradient. Because the pre-polarizing field $B_p$ was relatively low (10 mT), each measurement was repeated 200 times, and the results were averaged. The total acquisition time for the image in Fig. 7 was about 2 hours.

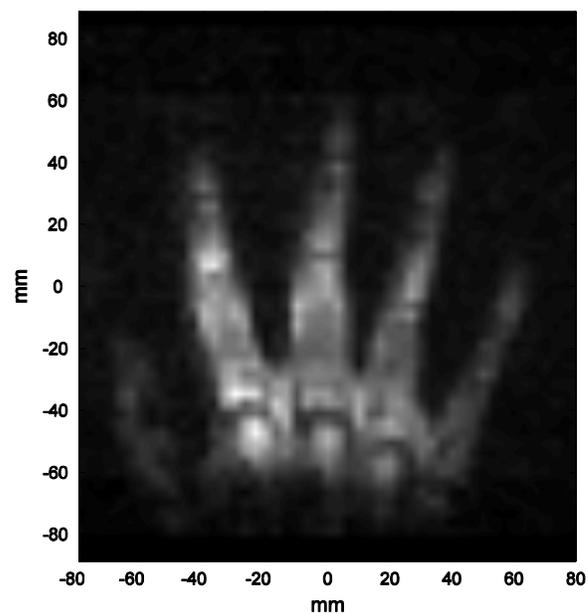

Fig. 7. Image of the human hand obtained as a combination of images from the seven individual channels.

Fig. 7 reveals some anatomical details of the hand such as the joints, represented by the dark elongated spots. The thumb, however, is barely visible, because the system's channels were placed above the other four fingers.

To improve the image, we applied the parallel imaging with localized sensitivities (PILS) technique [12]. Information about the channels' sensitivities was obtained by imaging a large (23 cm diameter) uniform phantom with the relaxation time $T_1$ ~ 120 ms. The imaging procedure was similar to the one described in Section II.C, but with 1 cm resolution along the phase-encoding direction. The combined sensitivity map after interpolation is shown in Fig. 8.

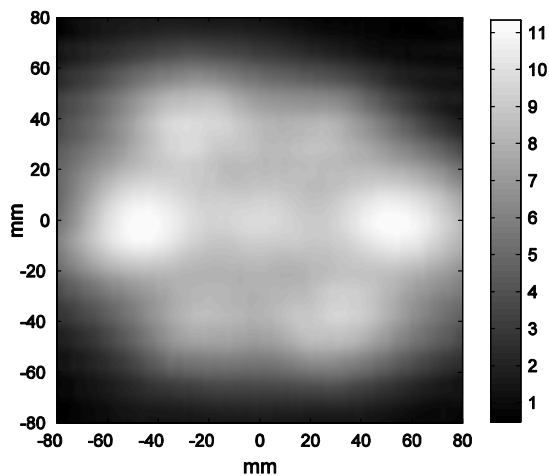

Fig. 8. Combined sensitivity map of the seven-channel system.

The sensitivity-corrected image is displayed in Fig. 9. It has more uniform intensity distribution and provides more details than the original image.

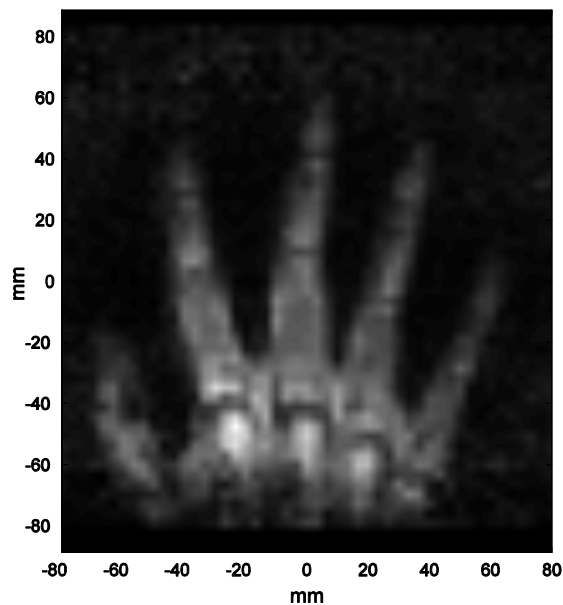

Fig. 9. Sensitivity-corrected image obtained using the PILS technique.

## IV. CONCLUSION

In this paper, we described the new seven-channel system capable of performing both MEG and ULF MRI. We also reported the first implementation of multi-channel ULF MRI and presented the multi-channel image of the human hand.

The first results of the in vivo ULF MRI were published recently by Mößle et al. [5]. These authors used a single-channel SQUID system for 3-D ULF MRI [4] to acquire images of human forearm and fingers [5]. The object width in their experiments was 40-60 mm. Our results, presented in this paper, show that the use of multiple channels allows imaging of larger objects: the human hand in our experiments was ~140 mm in size. We can also conclude that strong pre-polarizing field $B_p$ is essential for fast ULF MRI, because an increase in signal-to-noise ratio reduces an amount of time required for averaging. The images reported by Mößle et al. were acquired with pre-polarizing fields of 40-100 mT in 6 minutes [5]. In our experiments with the pre-polarizing field of 10 mT, the acquisition time was nearly 2 hours. A stronger $B_p$ field should reduce this time to an acceptable level.

The 2-D image of the hand, reported in this paper, demonstrates that our system can be successfully used for ULF MR imaging of large parts of the human body. Our next goal is to obtain 3-D images of parts of the human brain and use that spatial information to map MEG-localized sources of brain activity.

ACKNOWLEDGMENT

Technical development of the instrumentation used for this work was supported by the U.S. Department of Energy Office of Biological and Environmental Research, and by internal research funding from Los Alamos National Laboratory.